\documentclass[twocolumn,twoside,slac_two]{revtex4}
\usepackage{graphicx}
\usepackage{fancyhdr}
\begin{document}

\title{$D^0$ Mixing}

%

\author{B. Golob}
\affiliation{University of Ljubljana and Jozef Stefan
  Institute, Ljubljana, Slovenia}

\begin{abstract}
An overview of selected experimental results in the field of
$D^0$-$\bar{D}^0$ oscillations is presented. The average results for the
mixing parameters, $x=(0.89\pm{0.26\atop 0.27})\%$ and
$y=(0.75\pm{0.17\atop 0.18})\%$, exclude the no-mixing hypothesis at
the level of 6.7 standard deviations. No sign of $CP$ violation in the
$D^0$ system is observed. The measurements impose constraints on the
parameter space of many New Physics models.  
\end{abstract}

\maketitle

\thispagestyle{fancy}


\section{Introduction}

Last year, 31 years after the discovery of $D^0$ mesons, the first
evidence of a mixing phenomena in the system of neutral charm mesons has
been obtained \cite{belle_kk,babar_kpi}. Following this breakthrough
were several additional measurements enabling - through an averaging
procedure - a quite precise determination of the parameters governing
the mixing. The results presented in the paper follow from the data
collected by the two B-factories experiments, Belle and BaBar, from
the charm-factory experiment Cleo-c, as well as from the proton
collider experiment CDF. At the B-factories the 
cross-section for the continuum production of $c\bar{c}$ pairs is
around 1.3~nb, which with the integrated luminosity of KEKB amounts to
$10^9$ produced charmed hadron pairs. At CESR, the $\sim$800~pb$^{-1}$ data
sample corresponds to $2.8\times 10^6$ $D^0\bar{D}^0$ pairs produced in a
coherent $C=-1$ state. And while at the Tevatron the experimental
environment for the presented measurements is more difficult, the
cross-section for neutral charm meson production with a transverse momentum larger
than 5.5~GeV$/c$ yields a starting data
sample of $50\times 10^9$ $D^0$'s. The experiments thus provide a really
diverse experimental environment for successful studies in charmed hadron physics. 

The mixing, that is the transition of a neutral $D^0$ meson into its
antiparticle and vice-versa, appears as a consequence of states of
definite flavour ($D^0$, $\bar{D}^0$) being a linear superposition of
the mass eigenstates (states of simple exponential time evolution, $D_{1,2}$): 
\begin{equation}
|D_{1,2}\rangle=p|D^0\rangle\pm q |\overline{D}^0\rangle~~.
\label{eq0}
\end{equation}
It is governed by the lifetime of $D$ mesons, $\tau=1/\Gamma$, and by
the mixing parameters $x=(m_1-m_2)/\Gamma$ and
$y=(\Gamma_1-\Gamma_2)/2\Gamma$. $m_{1,2}$ and $\Gamma_{1,2}$ denote
the masses and widths of the mass eigenstates $D_1$ and $D_2$,
respectively. $\Gamma=(\Gamma_1+\Gamma_2)/2$ is the average decay width. The mixing
rate is small. The contribution of loop diagrams, successfully
describing the oscillations in other neutral meson systems, is
suppressed since the $D^0$ system
is the only neutral meson system with down-like quarks exchanged in
the loop. In this short distance calculation $x$ is negligibly small 
due to the small SU(3) flavor symmetry breaking
($m_s^2\approx m_{u,d}^2$) and elements of Cabibbo-Kobayashi-Maskawa
(CKM) matrix ($|V_{ub}|\approx 4\times
10^{-3}$) \cite{shipsey}. Long distance
contributions to the $D^0$-$\bar{D}^0$ transition are difficult to
calculate. Current theoretical estimates predict the mixing
parameters $|x|, |y|\le 10^{-2}$ \cite{falk,bigi1}. 

Violation of the $CP$ asymmetry ($CPV$) in the charm sector is expected to
be small. Since the processes with $D^0$ mesons involve mainly the
first two generations of quarks, for which the CKM elements are almost
real, the expected level of $CPV$ is ${\cal{O}}(10^{-3})$ which is below
the current experimental sensitivity. 

\section{Measurements}

Several methods and selection criteria are common to the presented
measurements. Tagging of the flavour of an initially produced $D^0$
meson is achieved by reconstruction of
decays $D^{\ast +}\to D^0\pi_s^+$ or $D^{\ast -}\to
\overline{D}^0\pi_s^-$. The charge of the characteristic low momentum
pion $\pi_s$ determines the tag. The
energy released in the $D^\ast$ decay, 
\begin{equation}
q=M(D^\ast)-M(D^0)-m_\pi~~,
\label{eq1}
\end{equation}
has a narrow peak for the signal events and thus
helps in rejecting the combinatorial background. 
Here, $M(X)$ is used to denote the invariant mass of the $X$
decay products, and $m_X$ stands for the nominal mass of $X$. $D^0$ 
mesons produced in $B$ decays have different decay time distribution
and kinematic properties than the mesons 
produced in fragmentation. In order to obtain a sample of neutral
mesons with uniform properties one selects $D^\ast$ mesons with
momentum above the kinematic limit for $B$ meson decays (B-factories)
or uses the impact parameter distribution to isolate primary charm mesons
(CDF). 

\subsection{Decays to CP eigenstates}

In the limit
of negligible $CPV$ the mass eigenstates $D_{1,2}$ are also
$CP$ eigenstates. In decays $D^0\to f_{CP}$ only the mass eigenstate 
component of $D^0$ with the $CP$ eigenvalue equal to the one of
$f_{CP}$ contributes. 
By measuring the lifetime of $D^0$ in decays
to $f_{CP}$ one thus determines the corresponding $1/\Gamma_1$ or
$1/\Gamma_2$. On the other
hand, flavour specific final states like $K^-\pi^+$ have a mixed $CP$
symmetry. The measured value of the effective lifetime in these decays 
corresponds to a mixture of $1/\Gamma_1$ and 
$1/\Gamma_2$. The relation between the two lifetimes can be written as 
\cite{bergman}
\begin{equation}
\tau(f_{CP})=\frac{\tau(D^0)}{1+\eta_f y_{CP}}~~, 
\label{eq2}
\end{equation}
where $\tau(f_{CP})$ and $\tau(D^0)$ are the lifetimes measured in
$D^0\to f_{CP}$ and $D^0\to K^-\pi^+$, respectively. 
$\eta_f=\pm 1$ denotes the $CP$ eigenvalue of $f_{CP}$. 
The relative difference of
the lifetimes is described by the parameter $y_{CP}$. 

$CP$-even final states $f_{CP}=K^+K^-, \pi^+\pi^-$ were used to
measure $y_{CP}$ \cite{belle_kk}.  
Expressed in terms of the mixing parameters, $y_{CP}$ reads
\cite{bergman} 
\begin{equation}
y_{CP}=y\cos{\phi}-\frac{1}{2}A_M\sin{\phi}~~, 
\label{eq3}
\end{equation}
with $A_M$ and $\phi$ describing the $CPV$ in mixing and interference
between mixing and decays, respectively. In case of no $CPV~(A_M, \phi
=0)$ and $y_{CP}=y$. 

\begin{figure}[h]
\centering
  \includegraphics[width=8cm]{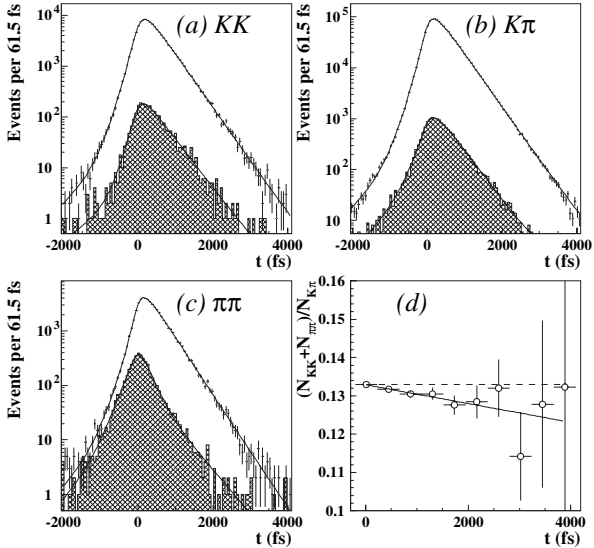}
  \caption{(a)-(c): Result of the simultaneous fit to decay time distributions in
    $D^0$ decays to $KK,~K\pi$ and $\pi\pi$ final states. The hatched areas
    represent the contribution of backgrounds. (d): Ratio of $D^0\to
    f_{CP}$ and $D^0\to K^-\pi^+$ decay time distributions. The slope
    visualizes the difference of effective lifetimes.}
  \label{fig1}
\end{figure}

Simultaneous fits to decay time distributions of selected $D^0\to K^+K^-, K^-\pi^+$
and $\pi^+\pi^-$ candidates were performed with $y_{CP}$ as a common free
parameter. The fit is presented in Fig. \ref{fig1}(a)-(c). 
The agreement of the fit function with the data
is excellent, $\chi^2/n.d.f=312/289$. The final value obtained is 
\begin{equation}
y_{CP}=(1.31\pm 0.32\pm 0.25)\%~~.
\label{eq4}
\end{equation}

The largest systematic uncertainties arise from the assumed resolution
function (common offset in individual decay modes), possible
deviations of 
acceptance dependence on decay
time from a constant (estimated by a fit to the generated $t$
distribution of reconstructed MC events) and variation of
selection criteria (effect estimated using high statistics MC
samples). 

The resulting $y_{CP}$ is more than 3
standard deviations above zero and hence represents clear evidence of
$D^0-\overline{D}^0$ mixing, regardless of possible $CPV$. The
difference of lifetimes is made 
visually observable by ploting the ratio of decay time distributions for decays to
$f_{CP}$ and $K^-\pi^+$ in Fig. \ref{fig1}(d). 

\begin{figure}[h]
\centering
  \includegraphics[width=6cm]{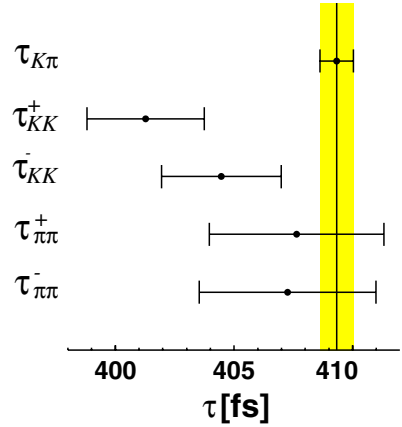}
  \caption{Lifetimes of $D^0$ mesons measured in decays to $K\pi$,
    $KK$ and $\pi\pi$ (separately for $D^0$ and $\bar{D}^0$ for the
    latter two modes) \cite{babar_kk}. The average value of the
    lifetime as measured in decays to CP eigenstates is lower than the
    one measured in flavour specific final state.}
  \label{fig2}
\end{figure}

Recently the BaBar collaboration performed a similar measurement \cite{babar_kk},
with results for individual lifetimes shown in Fig. \ref{fig2}. The
obtained value of the mixing parameter is 
\begin{equation}
y_{CP}=(1.24\pm 0.39\pm 0.13)\%~~.
\label{eq5}
\end{equation}

The two measurements make $y_{CP}$ the most precisely
measured individual mixing parameter in the $D^0$ system. 
 
Final states of definite $CP$ allow also a search for possible
$CPV$. Two methods, decay time dependent and time integrated, have been
exploited. For the former, the lifetime in $D^0\to f_{CP}$ is measured
separately for $D^0$ and $\bar{D}^0$ tagged events. The asymmetry is
\cite{bergman} 
\begin{eqnarray}
\nonumber
A_\Gamma&=&\frac{\tau(\overline{D}^0\to f_{CP})-\tau(D^0\to f_{CP})}
{\tau(\overline{D}^0\to f_{CP})+\tau(D^0\to f_{CP})}=\\
&=&\frac{1}{2}A_M y\cos{\phi} - x\sin{\phi}~~.
\label{eq6}
\end{eqnarray}
The values of $A_\Gamma$ measured by Belle \cite{belle_kk} and BaBar
\cite{babar_kk} are 
\begin{eqnarray}
\nonumber
A_\Gamma&=&(0.01\pm 0.30\pm 0.15)\%\\
A_\Gamma&=&(0.26\pm 0.36\pm 0.08)\%~~,
\label{eq7}
\end{eqnarray}
respectively, and show no sign of $CPV$ at the level of around 0.3\%. 

\begin{figure}[h]
\centering
  \includegraphics[width=8cm]{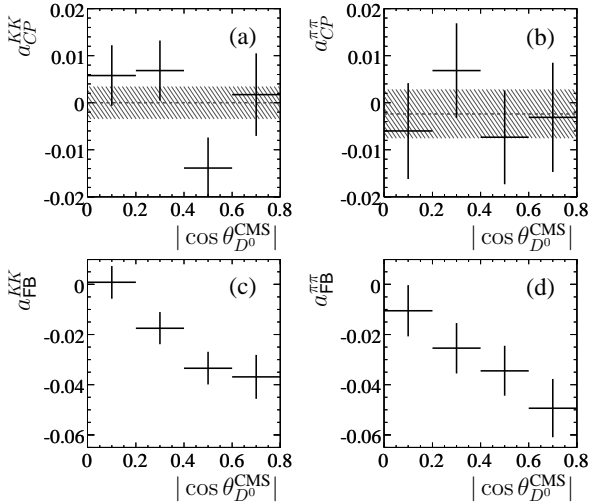}
  \caption{Time integrated $CP$ asymmetry ((a),(b)) and forward-backward asymmetry
    ((c),(d)) as a function of the $D$ meson polar angle
    \cite{babar_acp}.}
  \label{fig3}
\end{figure}

With the time integrated method one measures the asymmetry 
\begin{eqnarray}
\nonumber
A_{CP}&=&\frac{\Gamma({D}^0\to f_{CP})-\Gamma(\bar{D}^0\to f_{CP})}
{\Gamma({D}^0\to f_{CP})+\Gamma(\bar{D}^0\to f_{CP})}=\\
&=&a_{dec}^f+a_{mix}+a_{int}~~.
\label{eq8}
\end{eqnarray}
$A_{CP}$ receives contribution from all three types of $CPV$, direct,
$CPV$ in mixing and in the interference between decays with and without
the mixing. The latter two are independent of the final
state. Experimentaly the measured asymmetry must be corrected for
possible charge asymmetries in the detection of the slow pion as well as the
forward-backward asymmetry ($A_{FB}$) in the production of fermion pairs in
$e^+e^-$ collisions. The method of determination of $\pi_s$ correction factors
was developed in \cite{babar_acp} using the untagged $D^0\to K^-\pi^+$
decays. The forward-backward asymmetry is separated on the basis of
its symmetry properties as a function of the $D$ meson polar angle in the
center-of-mass system. Figure \ref{fig3} shows the measured $A_{CP}$
((a),(b)) and $A_{FB}$ ((c),(d)) as a
function of the $D$ meson polar angle \cite{babar_acp}. 
Averaging over the polar angle yields the value 
\begin{equation}
A_{CP}^{KK}=(0.00\pm 0.34\pm 0.13)\%~~. 
\label{eq9}
\end{equation}
Measurement of the Belle collaboration \cite{belle_acp}
\begin{equation}
A_{CP}^{KK}=(-0.43\pm 0.30\pm 0.11)\%~~. 
\label{eq10}
\end{equation}
is also consistent with no $CPV$. 

\subsection{Wrong-sign decays to hadronic final states}

Decays of $D$ mesons to two-body hadronic final states accessible to
both, $D^0$ and $\bar{D}^0$, have traditionally been used to search
for the mixing. Final state $K^+\pi^-$ can be reached through a doubly
Cabibbo supressed (DCS) $D^0$ decay as well as through the
$D^0\to\bar{D}^0$ mixing followed by a Cabibbo favoured (CF) decay. The
time evolution for these decays has three terms: 
\begin{equation}
|\langle K^+\pi^-|D^0(t)\rangle |^2\propto \bigl[R_D+\sqrt{R_D}y^\prime t+
 \frac{x^{\prime 2}+y^{\prime 2}}{4} t^2\bigr]e^{-t}~~.
\label{eq11}
\end{equation}
The first one is due to DCS decays, the third one due to the mixing, and
the middle term represents the interference of the two contributions. $R_D$
is the Cabibbo suppression factor relative to CF decays. $x^\prime$
and $y^\prime$ are the mixing parameters, rotated by a strong phase
difference between DCS and CF decays, $x^\prime =x \cos\delta + y\sin\delta$ and 
$y^\prime=y\cos\delta - x\sin\delta$. The dimensionless time $t$ is
measured in units of $\tau(D^0)$. 

\begin{figure}[h]
\centering
  \includegraphics[width=6cm]{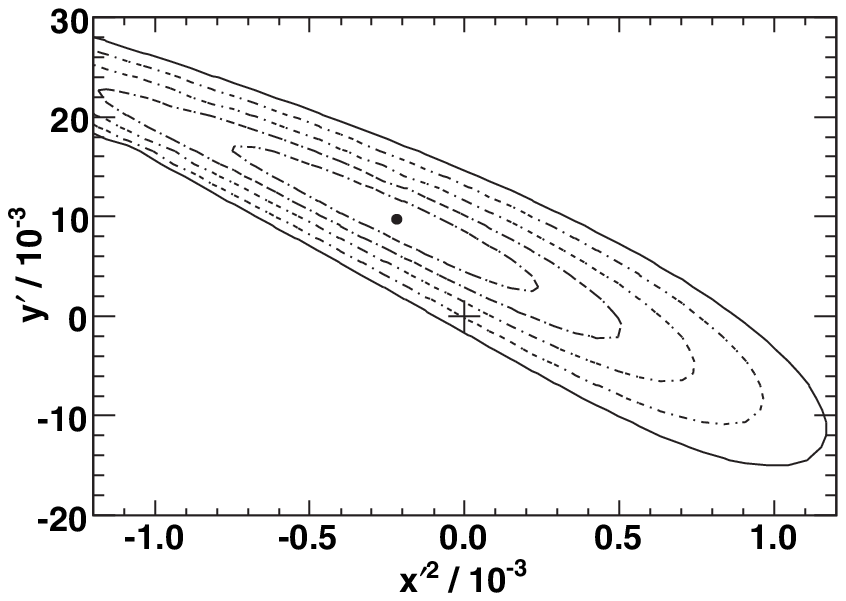}
\includegraphics[width=4cm]{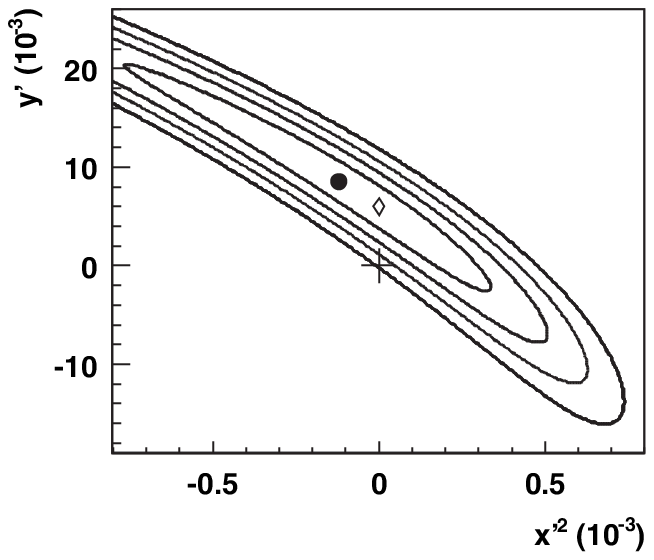}\includegraphics[width=4cm]{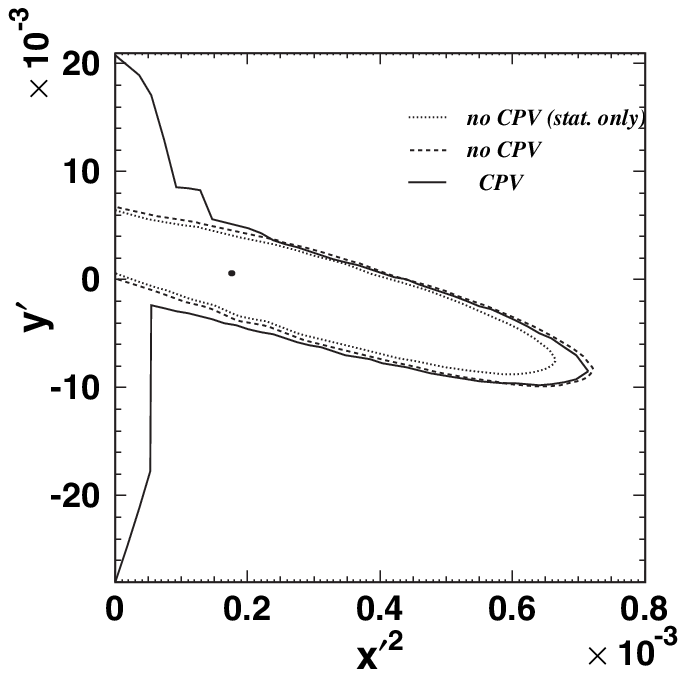}
  \caption{Top: 1 - 5 $\sigma$ likelihood contours for parameters $x^{\prime 2}$ and
    $y^\prime$ obtained from a fit to the decay time distribution of
    $D^0\to K^+\pi^-$ decays \cite{babar_kpi}. Bottom left: Bayesian
    probability contours in $x^{\prime 2},y^\prime$ plane
    corresponding to 1 - 4 $\sigma$ as obtained in
    \cite{cdf_kpi}. Bottom right: 95\% C.L. contours of $x^{\prime
    2},y^\prime$ using
    Feldman-Cousins approach from \cite{belle_kpi}.}
  \label{fig4}
\end{figure}

The BaBar collaboration obtained the
first evidence for $D^0$ mixing by performing the decay time study of
wrong-charge decays $D^{\ast +}\to D^0\pi_s^+$, $D^0\to K^+\pi^-$ to
separate the DCS and the mixing contribution. The 
parameters obtained from the fit are presented in terms of likelihood
contours in Fig. \ref{fig4} (top). The central value lies slightly in the
non-physical region ($x^{\prime 2}<0$) and the no-mixing point $(x^{\prime
    2}=0,y^\prime=0)$ is excluded at the level corresponding to 3.9
standard deviations. Recently CDF collaboration obtained the result
of similar significance \cite{cdf_kpi} shown in Fig. \ref{fig4}
(bottom left). Result form the Belle collaboration \cite{belle_kpi} takes into account
the presence of a physical boundary and is presented in
Fig. \ref{fig4} (bottom right) as a 95\% C.L. contour calculated using
the Feldman-Cousins method. 

\subsection{Time dependent Dalitz analyses}

Several intermediate resonances can contribute to a hadronic multi-body
final state. In a specific decay channel $D^0\to K_S\pi^+\pi^-$, 
recently analyzed by Belle \cite{belle_kspipi},  
contributions from CF decays (e.g. $D^0\to K^{\ast
  -}\pi^+$), DCS decays (e.g. $D^0\to K^{\ast
  +}\pi^-$) and decays to $CP$ eigenstates (e.g. $D^0\to \rho^0 K_S$)
are present. Individual contributions can be identified by analyzing
the Dalitz distribution of the decay. Moreover, for a self-conjugated
final state these different types of decays interfere and it is possible to
determine their relative phases (unlike in the case of $D^0\to K^+\pi^-$ decays). 
Since these types of intermediate states also exhibit 
a specific time evolution one can determine directly the mixing
parameters $x$ and $y$ by studying the time evolution of the Dalitz
distribution. 

\begin{figure}[h]
  \includegraphics[width=6cm]{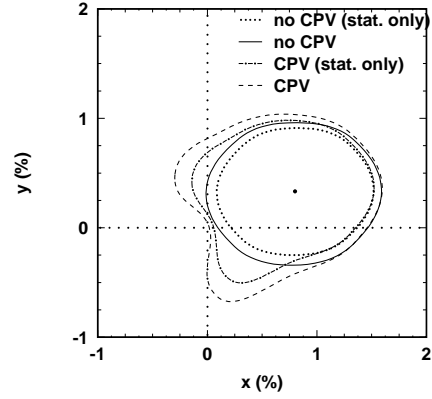}
  \caption{95\% C.L. region for parameters $x$ and $y$ as obtained in 
$D^0\to K_S\pi^+\pi^-$ decays \cite{belle_kspipi}.}
  \label{fig5}
\end{figure}

The signal p.d.f. for a simultaneous fit to the Dalitz and decay-time
distribution is 
\begin{eqnarray}
\nonumber
&&{\cal{M}}(m_-^2,m_+^2,t)=\langle K_S\pi^+\pi^-|D^0(t)\rangle=\\
\nonumber
&=&\frac{1}{2}{\cal{A}}(m_-^2,m_+^2)\bigl[e^{-i\lambda_1
      t}+e^{-i\lambda_2 t}\bigr]+\\
&+&\frac{1}{2}\overline{\cal{A}}(m_-^2,m_+^2)\bigl[e^{-i\lambda_1 t}-
  e^{-i\lambda_2 t}\bigr]~~.
\label{eq12}
\end{eqnarray}
The matrix element is composed of an instantaneous amplitude for $D^0$
decay, ${\cal{A}}(m_-^2,m_+^2)$, and an amplitude for the $\bar{D}^0$
decay, $\bar{\cal{A}}(m_-^2,m_+^2)$, arising due to a possibility
of mixing. They both depend on the Dalitz variables $m_-^2=M^2(K_S\pi^-)$ and
$m_+^2=M^2(K_S\pi^+)$. The dependence on the mixing parameters is hidden in 
$\lambda_{1,2}=m_{1,2}-i\Gamma_{1,2}/2$. If $CPV$ is neglected the
amplitude for $\bar{D}^0$ tagged decays is 
$\bar{\cal{M}}(m_+^2,m_-^2,t)={\cal{M}}(m_-^2,m_+^2,t)$. 
Amplitudes for $D$ decays are parametrized in the isobar model as a
sum of 18 Breit-Wigner resonances and a constant non-resonant term. The
result of the fit in terms of mixing parameters is presented in
Fig. \ref{fig5}. 

Numerically, the fit which allows for the $CPV$ results in 
\begin{eqnarray}
\nonumber
x&=&(0.80\pm 0.29\pm {0.13\atop 0.16})\%\\
\nonumber
y&=&(0.33\pm 0.24\pm {0.10\atop 0.14})\%\\
\nonumber
|q/p|&=&0.86\pm {0.30\atop 0.29}\pm {0.10\atop 0.09})\%\\
\phi&=&(-0.24\pm {0.28\atop 0.30}\pm 0.09)~{\rm rad}~~.
\label{eq13}
\end{eqnarray}
The measurement represents the most accurate determination of $x$. The
$CP$ violating parameters $|q/p|$ and $\phi=arg(q/p)$ are consistent
with no $CP$ violation. 

BaBar has performed the time dependent Dalitz
analysis of $D^0\to K^+\pi^-\pi^0$ decays \cite{babar_kpipi0}. The final state is flavour
specific and hence the wrong-sign decays again receive contribution
from mixing and DCS process. The combined Dalitz-decay-time signal
distribution has the form 
\begin{eqnarray}
\nonumber
&&|\langle
K^+\pi^+\pi^-|D^0(t)\rangle|^2\propto\bigl[|A_{\bar{f}}|^2+\\
\nonumber
&+&|\bar{A}_{\bar{f}}||A_{\bar{f}}|(y^{\prime\prime}\cos\delta_f - 
x^{\prime\prime}\sin\delta_f)t+\\
&+&|\bar{A}_{\bar{f}}|^2\frac{x^{\prime\prime 2}+y^{\prime\prime
    2}}{4} t^2 \bigr] e^{-t}~~.
\label{eq14} 
\end{eqnarray}
$A_{\bar{f}}$ (depending on the Dalitz variables $M^2(K^+\pi^-)$ and
$M^2(K^+\pi^0)$) is the amplitude for $D^0\to K^+\pi^-\pi^0$
decays determined from the fit to the Dalitz distribution of
wrong-sign decays. The amplitude for $\bar{D}^0$
decays, $\bar{A}_{\bar{f}}$, is fixed to the values obtained in the fit to the time
integrated Dalitz distribution of right-sign $D^0\to K^-\pi^+\pi^0$
decays. The relative phase $\delta_f$ also depends on
$M^2(K^+\pi^-),~M^2(K^+\pi^0)$, and is determined from the fit to the wrong- and
right-sign Dalitz distributions. Parameters $x^{\prime\prime}$ and $y^{\prime\prime}$ are, 
similar as in the case of $D^0\to K^+\pi^-$ decays, a rotated mixing
parameters $x$ and $y$, now by an unknown strong phase shift $\delta_{K\pi\pi^0}$
between two points in phase spaces of DCS and CF decays to $K^-\pi^+\pi^0$. 

A fit to the time evolution of the wrong-sign Dalitz distribution
results in 
\begin{eqnarray}
\nonumber
x^{\prime\prime}&=&(2.61\pm {0.57\atop 0.67}\pm 0.39)\%\\
y^{\prime\prime}&=&(-0.06\pm {0.55\atop 0.64}\pm 0.34)\%~~.
\label{eq15}
\end{eqnarray}

\subsection{$\psi(3770)\to D^0\bar{D}^0$}

Pairs of neutral $D$ mesons are produced at the threshold in a coherent
$C=-1$ state. With the Cleo-c detector one exploits this quantum
coherence to determine several parameters related to the mixing. The
effective branching ratios for $\psi(3770)\to D^0\bar{D}^0\to f_1f_2$
are modified with respect to the values measured from uncorrelated
$D^0$ meson decays. The branching fractions can be obtained by
integration of the time dependence for 
$\psi(3770)\to D^0\bar{D}^0\to f_1f_2$ \cite{PDG}, 
\begin{eqnarray}
\nonumber
&&\frac{d\Gamma(\psi(3770)\to f_1f_2)}{d\Delta t}\propto 
\bigl(|a_+|^2+|a_-|^2\bigr) \cosh{y\Delta t}+\\
\nonumber
&+&\bigl(|a_+|^2-|a_-|^2\bigr) \cos{x\Delta t} -\\
&-&2\Re(a_+^\ast a_-)\sinh{y\Delta t}+2\Im(a_+^\ast a_-)\sin{x\Delta
  t}~~, 
\label{eq16}
\end{eqnarray}
where $\Delta t$ is the time interval between two $D$ mesons decays,
and amplitudes $a_+$ and $a_-$ are in the limit of no $CPV$ defined as 
$a_+\equiv \bar{A}_{f_1}A_{f_2}-A_{f_1}\bar{A}_{f_2}$, $a_-\equiv
A_{f_1}A_{f_2}-\bar{A}_{f_1}\bar{A}_{f_2}$. 
The branching fraction for decays to the final state
    composed of a CP-even state ($S_+$) and semileptonic state
    ($e^-X$), for example, is found to be  $Br(\psi(3770)\to
    D^0\bar{D}^0\to S_+e^-X)\approx Br(D^0\to
    S_+)Br(\bar{D}^0\to
    e^-X)(1-y)$. In a similar manner other double-tagged (both $f_1$
    and $f_2$ reconstructed) branching fractions depend on the parameters $y,~R_M=(x^2+y^2)/2$ and
    $\sqrt{R_D}\cos\delta$ \cite{cleo_c}. Single-tagged (only a single
    final state reconstructed) branching fraction ($Br(D^0\to
    S_+)$ and $Br(\bar{D}^0\to
    e^-X)$ in the above example) remain unchanged
    compared to the uncorrelated measurements.  
 
\begin{figure}[h]
  \includegraphics[width=8cm]{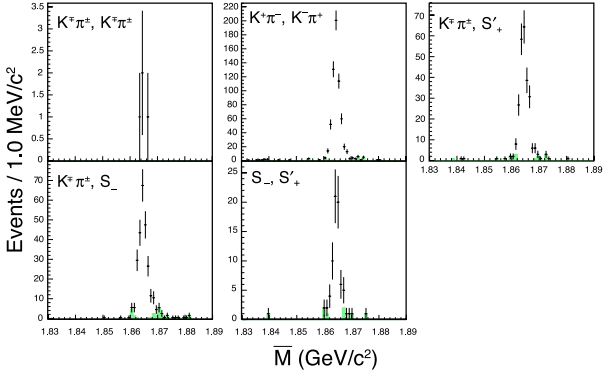}
  \caption{Fully reconstructed $\psi(3770)\to
    D^0\bar{D}^0\to f_1f_2$ decays \cite{cleo_c}. The hatched area
    represents events outside the signal region.}
  \label{fig6}
\end{figure}

The measurements consist of determination of a set of double- and single-tagged
branching ratios. Examples of reconstructed double-tagged signal
yields are shown in Fig. \ref{fig6}. 
Parameters mentioned above are obtained from a simultaneous fit to the measured single- and double-tagged branching
ratios (including additional world-average information on individual
branching fractions). The result is \cite{cleo_c} 
\begin{eqnarray}
\nonumber
y&=&(-5.2\pm 6.0\pm 1.7)\%\\
\nonumber
R_M&=&(2.0\pm 1.2\pm 1.2)\times 10^{-3}\\
\sqrt{R_D}\cos\delta &=&0.089\pm 0.036\pm 0.009~~.
\label{eq17}
\end{eqnarray}
This is the only direct measurement of the strong phase difference
$\delta$ between DCS and CF decays $D^0\to K^\pm\pi^\mp$. 

\subsection{Average of results}

\begin{figure}[h]
  \includegraphics[width=8cm]{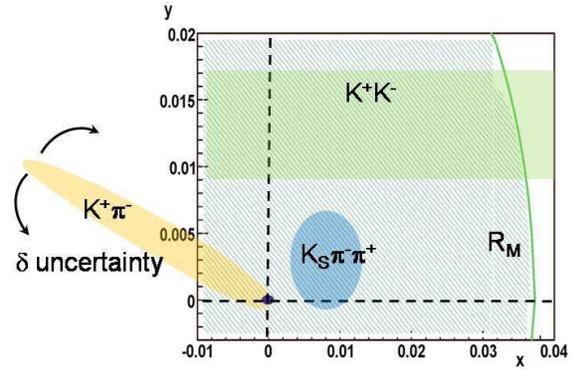}
  \caption{Illustration of constraints in the $(x,y)$ plane
    imposed by measurements in individual decay modes. $R_M$
    corresponds to measurements of semileptonic $D^0$ decays.}
  \label{fig7}
\end{figure}

Measurements presented in the previous sections as well as others
constrain the possible values of the $D^0$ mixing and $CPV$ parameters
in a way specific to the decay mode under consideration. A schematic
view of constraints on $x$ and $y$ posed by some of the measurements
is presented in Fig. \ref{fig7}. 
Heavy flavour averaging group \cite{hfag} performs a $\chi^2$ fit
to the $D^0$ mixing and $CPV$ related
observables. Correlations among the measured quantities are provided
by the experiments and included in the fit. The average values from
the fit in 
which $CPV$ is
allowed for are presented in Tab. \ref{tab1}. 
\begin{table}[h]
\begin{center}
\caption{World average values of mixing and $CP$ violation parameters
  \cite{hfag}.}
\begin{tabular}{|c|c|}
\hline \textbf{Parameter} & \textbf{Value} 
\\ \hline  
$x~[\%]$& $0.89\pm{0.26\atop 0.27}$ \\
$y~[\%]$& $0.75\pm{0.17\atop 0.18}$ \\
$\delta~[^\circ]$& $21.9\pm{11.3\atop 12.4}$ \\
$R_D~[\%]$& $0.3348\pm 0.0086$\\
$A_D~[\%]$& $-2.0\pm 2.4$ \\
$|q/p|$& $0.87\pm {0.18\atop 0.15}$\\
$\phi~[^\circ]$& $-9.1\pm {8.1\atop 7.8}$ \\
$\delta_{K\pi\pi^0} [^\circ]$& $33.0\pm {25.9\atop 26.6}$\\
\hline
\end{tabular}
\label{tab1}
\end{center}
\end{table}

The no-mixing scenario $(x,y)=(0,0)$ is excluded by the world average
of results at the level corresponding to almost 7 standard
deviations. Since positive values of $x$ and $y$ are preferred the
almost $CP$ even state of neutral $D$ mesons seems to be shorter-lived
and heavier. 

\section{Prospects and summary}

\begin{figure}[h]
  \includegraphics[width=8cm]{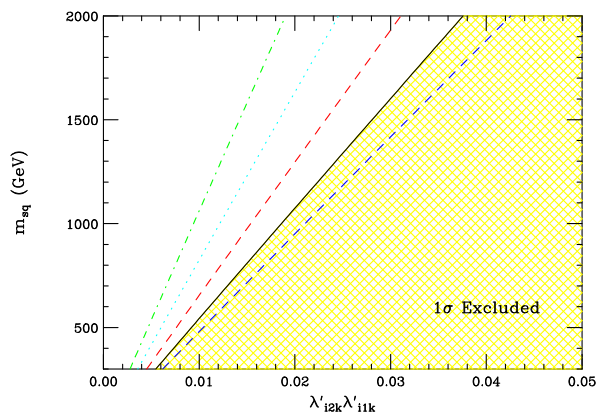}
  \caption{Constraints in squark mass - R-parity violating couplings
    plane 
    arising from the established value of $x$. The hatched area
    represents the excluded region from $x<1.2~\%$. The lines
    represent excluded regions for $x$ below 1.5\% - 0.3\% \cite{golowich}.}
  \label{fig8}
\end{figure}

A rich harvest of results in the field of $D^0$ mixing, arising from
a range of various experiments, only in the last two
years established the oscillations of this neutral mesons with a
significance of around seven standard deviations. The measured mixing
parameters $x$ and $y$ of ${\cal{O}}(10^{-2})$ are at the upper edge of
the values that can be accomodated within the SM. The results impose
severe constraints on the parameter space of a wide range of New Physics models
\cite{golowich}. To illustrate this, Fig. \ref{fig8} shows the
constraints on the squark mass - coupling constants plane arising from
the established value of $x$ in the supersymmetric 
R-parity violating models. 

Several measurements focus on the search for $CP$ violation in the
$D^0$ system where a positive signal at the current level of
sensitivity would represent a clear indication of New Physics
processes. At the moment there is no such hint, and $CPV$ is in
several processes found to be below the 0.3\% level. 

In the near future the major experimental task in the field is to
measure $x$ with a higher accuracy and to further limit the range of
the $CP$ violation in various decay modes. Recent experimental results
might trigger further theoretical efforts in the SM predictions for the
mixing parameters, although at the moment more precise calculations
appear to be difficult. 

A more or less educated guess tells that with a modest integrated 
luminosity of 5~ab$^{-1}$ collected by the future Super-B factory it
would be possible to measure $x$ and $y$ with a precission of 0.1\% -
0.15\%. Furthermore, the $CPV$ parameters $|q/p|$ and $\phi$ could be
measured to around $\pm 0.1$, enabling tests of the $CP$ asymmetries
at around $10^{-3}$ level. 


\bigskip 

\end{document}